\documentclass[%
 reprint,
 amsmath,amssymb,
aip,jcp
]{revtex4-2}

\usepackage{graphicx}% Include figure files
\usepackage{dcolumn}% Align table columns on decimal point
\usepackage{bm}% bold math
\UseRawInputEncoding
\begin{document}

\preprint{APS/123-QED}

% \title{Submicron-scale Control of Local Phase in Er:TiO$_2$ Films by Laser Annealing}
% \title{Quasi-deterministic Placement of Er Emitters in Thin Film TiO$_2$ through Local Phase Control with Diffraction-limited Spatial Resolution}
\title{Quasi-deterministic Localization of Er Emitters in Thin Film TiO$_2$ through Submicron-scale Crystalline Phase Control}

\author{Sean E. Sullivan}%
\thanks{These authors contributed equally.}
\email[Electronic mail: ]{sean@memq.tech}
\affiliation{%
 memQ, Inc. Chicago, IL 60615, United States
}%

 % \altaffiliation{These authors contributed equally.}%Lines break automatically or can be forced with \\
\author{Jonghoon Ahn}%
\thanks{These authors contributed equally.}
\email[Electronic mail: ]{sean@memq.tech}
 \affiliation{
Center for Molecular Engineering, Argonne National Laboratory, Lemont, IL 60439, United States
}%
 \affiliation{
Materials Science Division, Argonne National Laboratory, Lemont, IL 60439, United States
}%

 % \altaffiliation{These authors contributed equally.}%Lines break automatically or can be forced with \\

\author{Tao Zhou}
 \affiliation{
Center for Nanoscale Materials, Argonne National Laboratory, Lemont, IL 60439, United States
}%

% \collaboration{MUSO Collaboration}%\noaffiliation
\author{Preetha Saha}
\affiliation{%
 memQ, Inc. Chicago, IL 60615, United States
}%c. Chicago, IL

\author{Martin V. Holt}
 \affiliation{
Center for Nanoscale Materials, Argonne National Laboratory, Lemont, IL 60439, United States
}%
% \author{David D. Awschalom}
%  \affiliation{
% Pritzker School of Molecular Engineering, University of Chicago
% }%
%  \affiliation{
% Center for Molecular Engineering and Materials Science Division, Argonne National Laboratory
% }%

\author{Supratik Guha}
 \affiliation{
Pritzker School of Molecular Engineering, University of Chicago, Chicago, IL 60637, United States
}%
 \affiliation{
Materials Science Division, Argonne National Laboratory, Lemont, IL 60439, United States
}%
\author{F. J. Heremans}
 \affiliation{
Center for Molecular Engineering, Argonne National Laboratory, Lemont, IL 60439, United States
}%
 \affiliation{
Materials Science Division, Argonne National Laboratory, Lemont, IL 60439, United States
}%
\affiliation{
Pritzker School of Molecular Engineering, University of Chicago, Chicago, IL 60637, United States
}%

\author{Manish Kumar Singh}
% \email{
%  manish@memq.tech
% }%
\affiliation{%
 memQ, Inc. Chicago, IL 60615, United States
}%
% \collaboration{CLEO Collaboration}%\noaffiliation

\date{\today}% It is always \today, today,

\begin{abstract}
With their shielded 4f orbitals, rare-earth ions (REIs) offer optical and electron spin transitions with good coherence properties even when embedded in a host crystal matrix, highlighting their utility as promising quantum emitters and memories for quantum information processing. Among REIs, trivalent erbium (Er$^{3+}$) uniquely has an optical transition in the telecom C-band, ideal for transmission over optical fibers, and making it well-suited for applications in quantum communication. The deployment of Er$^{3+}$ emitters into a thin film TiO$_2$ platform has been a promising step towards scalable integration; however, like many solid-state systems, the deterministic spatial placement of quantum emitters remains an open challenge. We investigate laser annealing as a means to locally tune the optical resonance of Er$^{3+}$ emitters in TiO$_2$ thin films on Si. Using both nanoscale X-ray diffraction measurements and cryogenic photoluminescence spectroscopy, we show that tightly focused below-gap laser annealing can induce anatase to rutile phase transitions in a nearly diffraction-limited area of the films and improve local crystallinity through grain growth. As a percentage of the Er:TiO$_2$ is converted to rutile, the Er$^{3+}$ optical transition blueshifts by 13 nm. We explore the effects of changing laser annealing time and show that the amount of optically active Er:rutile increases linearly with laser power. We additionally demonstrate local phase conversion on microfabricated Si structures, which holds significance for quantum photonics.

% \begin{description}
% Secondary publications and information retrieval purposes.
% You may use the \texttt{description} environment to structure your abstract;
% use the optional argument of the \verb+\item+ command to give the category of each item. 
% \end{description}
\end{abstract}

\maketitle

%\tableofcontents

\section{\label{sec:intro}Introduction}
% \begin{itemize}
%   \item Er and REIs as qubits for quantum information science, quantum communication
%   \item Properties of different host materials for Er (Jeff Thompson's work)
%   \item Integration in thin film platform (cite our prior work). Different phases of TiO2
%   \item TiO2 as a well-studied material; demonstration of local phase control using laser annealing
%   \item In this work, we demonstrate control of the phase of Er-doped TiO2 thin films on silicon with diffraction-limited resolution. We confirm the phase change with low-temperature photoluminescence measurements, as well as nanoscale X-ray diffraction measurements.  
% \end{itemize}

Solid-state quantum emitters in wide-bandgap semiconductors offer atom-like level structures, making them candidate optically addressable spin qubits. \cite{wolfowicz_quantum_2021} Among these quantum emitters, rare-earth ions (REIs), are promising for applications in quantum information processing owing to their long optical\cite{boettger_material_2003} and spin coherence times.\cite{rancic_coherence_2018,le_dantec_twenty-threemillisecond_2021,gupta_robust_2023} Trivalent erbium (Er$^{3+}$), in particular, offers an optical transition in the telecom C-band (near 1.5 $\mu$m), which is well suited for optical fiber-based quantum communication.\cite{pettit_perspective_2023} In addition, integration of Er-doped crystals with nanophotonic cavities has enabled emission of both single\cite{dibos_atomic_2018} and indistinguishable\cite{ourari_indistinguishable_2023} photons, and single-shot optical readout of hyperfine spin states.\cite{raha_optical_2020,chen_parallel_2020} In contrast to trapped neutral atoms or ions, the properties of solid-state spin qubits are intimately tied to their environment, i.e., the host crystal in which they are embedded. While the local crystal environment can be a source of noise or relaxation from charge carriers or phonons, it also offers a potential phase space for tuning the properties of the quantum emitter through materials engineering. Although the 4f-4f optical transitions of REIs are somewhat shielded from nearby sources of noise, the optical lifetimes, optical coherence, and transition energies are nevertheless influenced by the local crystal environment.\cite{ortu_simultaneous_2018} For Er$^{3+}$, several suitable oxide host materials have been identified that minimize the REI’s first-order sensitivity to local electric fields by virtue of inversion-symmetric crystal sites. \cite{stevenson_erbium-implanted_2022} Furthermore, the spin coherence properties of these REIs can be enhanced by selecting host crystals with a naturally low abundance of isotopes that have nuclear spins, which can introduce phase noise through magnetic dipolar interactions. \cite{ferrenti_identifying_2020, le_dantec_twenty-threemillisecond_2021, kanai_generalized_2022} One such promising host material on both accounts is TiO$_2$, which has been explored as a host material with very low inhomogeneous broadening ($<$ 460 MHz) in its bulk crystalline form. \cite{phenicie_narrow_2019} 

Importantly, high quality TiO$_2$ films can readily be synthesized and integrated into standard CMOS process flows, \cite{djordjevic_cmos-compatible_2013} enabling a pathway towards scalable fabrication. Given its standard thin film integration, several recent works have investigated TiO$_2$ thin films as a host for Er quantum emitters in rutile, anatase, and mixed-phase polycrystalline forms. \cite{singh_development_2022,dibos_purcell_2022,shin_er-doped_2022} These works showed that optical properties of the TiO$_2$ films could be tuned through engineering of the thin film stack by delta doping, epitaxial growth, or by adjusting the growth temperature to select either the lower temperature anatase crystal phase or the higher temperature rutile phase in aggregate. As one of the most heavily studied metal oxide materials for photoelectrochemistry, the optical properties of TiO$_2$ nanoparticles and films doped with Er have also been investigated for applications in photocatalysis and photodetection, where high concentrations ($\lesssim$1 at\%) of Er act as up-conversion sensitizers in the visible range. \cite{mondal_investigation_2018, rao_defect_2019,kot_influence_2023} 

In the anatase phase, the optical transition between the lowest crystal field-split levels in the ground and first excited states (Z$_1$-Y$_1$) emits at a wavelength near 1533 nm. \cite{johannsen_influence_2016,shin_er-doped_2022} Meanwhile, in the rutile phase, the Z$_1$-Y$_1$ transition is near 1520 nm. \cite{phenicie_narrow_2019} This considerable difference in the transition energies between the two phases yields opportunities for spectrally isolating quantum emitters of interest. In parallel, prior efforts have demonstrated the ability to locally adjust the crystalline phase of TiO$_2$ films and nanoparticles by laser annealing, showing conversion from amorphous films to a crystalline phase or from anatase to rutile as detected by micro-Raman spectroscopy\cite{vasquez_laser-induced_2015,benavides_laser_2018,ahmed_localized_2021} or transmission electron microscopy.\cite{abbasi_situ_2023} As with other solid-state quantum emitters, such as those based on crystalline defects, a key challenge for scalable quantum technologies is the deterministic spatial placement of the emitter within a desired crystal volume. Laser writing has been used as a localization technique for generating vacancy complex-based quantum emitters in diamond\cite{chen_laser_2019} and silicon carbide.\cite{chen_laser_2019-1} In a similar vein, we utilize micro-focused laser annealing as a means toward localization of Er$^{3+}$ emitters in TiO$_2$ thin films. Here, we demonstrate localized anatase to rutile conversion of moderately Er-doped TiO$_2$ films with near diffraction-limited spatial resolution. We use cryogenic photoluminescence mapping and nanoprobe synchrotron-based X-ray diffraction measurements to confirm the phase transition and to measure rutile grain size. 

\section{\label{sec:main}Main}
\begin{figure}[htbp!]
\centering
\includegraphics[width=0.5\textwidth]
{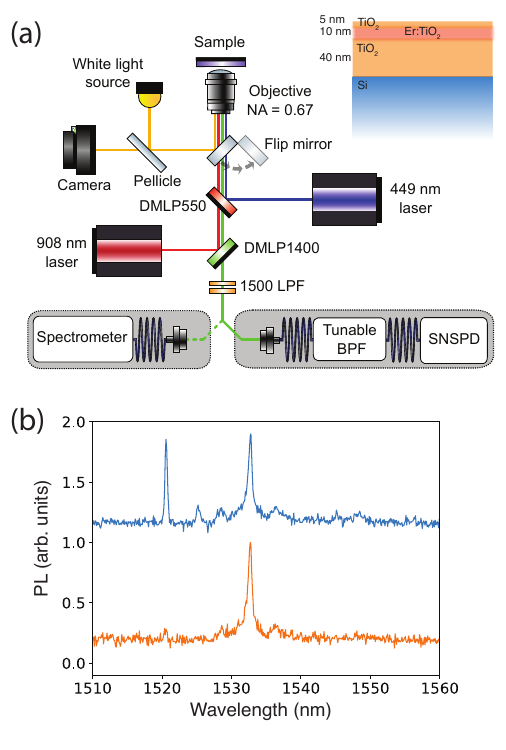}
\caption{(a) A simplified schematic of the experimental setup (DMLP: dichroic mirror long pass, LPF: long pass filter, BPF: band pass filter, SNSPD: superconducting nanowire single-photon detector). A blue laser ($\lambda = 449$ nm) is used for the laser anneal of the Er-doped TiO$_2$ films. An off-resonant laser ($\lambda = 908$ nm) is used for the optical excitation for the cryogenic photoluminescence measurements. The Er emission collected by the objective passes through dichroic mirrors and filters and is detected by the spectrometer or the SNSPD. The inset shows a schematic of the Er-doped TiO$_2$ film stack. (b) Representative photoluminescence spectra of Er:TiO$_2$ before laser annealing (orange) and after annealing (blue). Measurements taken at T=3.6 K.}
\label{fig1} 
\end{figure}

Polycrystalline thin films of TiO$_2$ were grown on silicon substrates in a molecular beam deposition chamber. The growth method and phase control via growth temperature are described in more detail elsewhere.\cite{singh_development_2022} In this work, all measurements were performed on a 55 nm-thick polycrystalline anatase film grown on 3-inch Si(100), which comprised three layers (40 nm, 10 nm, and 5 nm from bottom to top) with the middle 10 nm doped with erbium at a $\sim$500 parts per million (ppm) concentration (Fig. \ref{fig1}(a) inset). This wafer was subsequently diced into chips for different annealing experiments. The particular film stack was chosen to maximize the observed brightness of the Er$^{3+}$ photoluminescence in this experiment while maintaining a thin profile with an eye toward integrated quantum photonic applications. \cite{dibos_purcell_2022}

The films were loaded into an optical cryostat integrated with a custom confocal microscope setup as shown in Fig. \ref{fig1}(a). During the laser annealing process, the films were kept at room temperature and under ambient atmospheric conditions while a variable power $\lambda = 449$ nm continuous wave (CW) laser was tightly focused through an air objective lens with an N.A. of 0.67. The objective lens was mounted on a three-dimensional periscope stage, which allowed the control of the position and focus of the laser beam with respect to the sample films. The laser anneal was performed in sets of arrays of 12 spots with 5 $\mu$m spacings, wherein among the different sets, the effective $\lambda = 449$ nm laser power was varied from below 100 mW to approximately 185 mW (at sample position) and the duration of the laser anneal was varied from 1 to 100 seconds. After the laser annealing step, the films were quickly evaluated for phase change as-is, or they underwent a post-laser-annealing process in a tube furnace in a 20\% O$_2$/balance N$_2$ environment heated to either 400 $^\circ$C or 500 $^\circ$C for one hour. A quick verification of phase change could be confirmed by measuring the photoluminescence (PL) spectra of the films (Fig. \ref{fig1}(b)). 

\begin{figure*}[htbp!]
\centering
\includegraphics[width=1\textwidth]{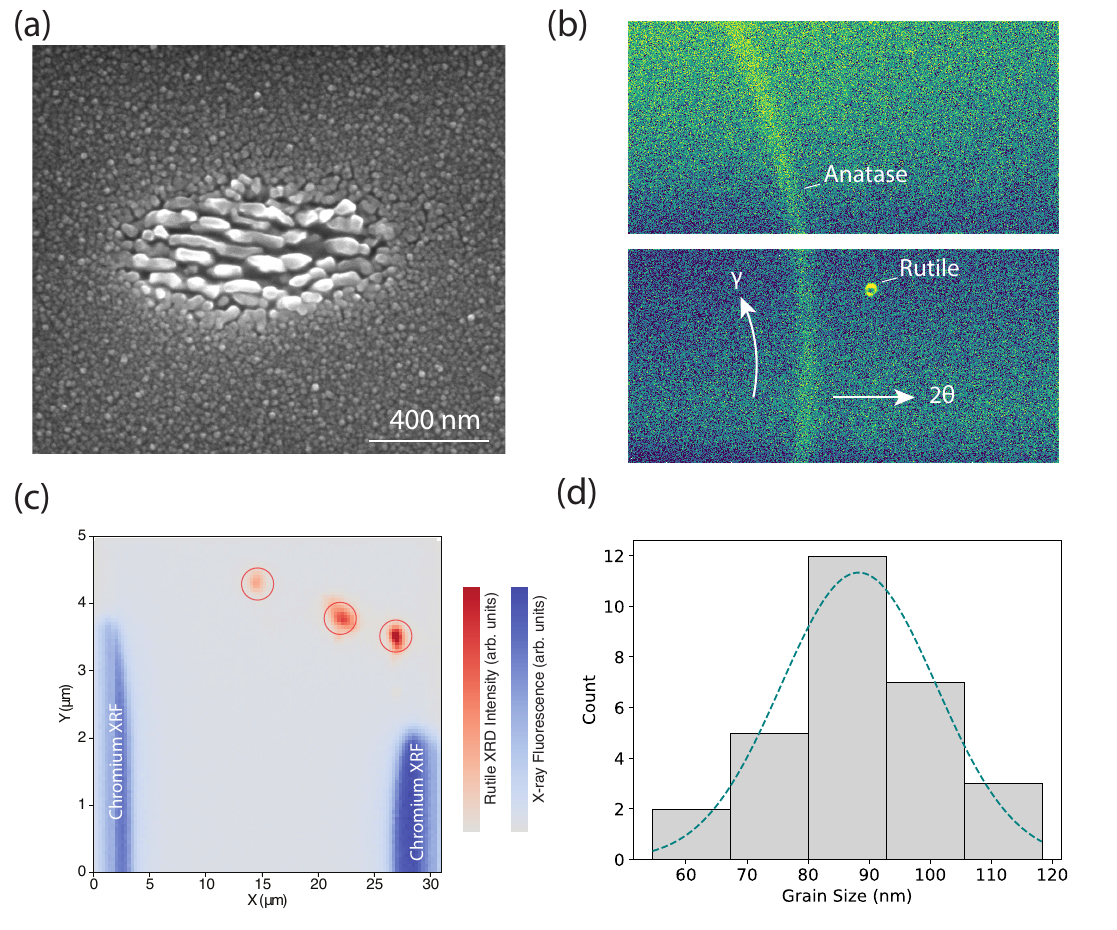}
\caption{ (a) Scanning electron micrograph of a single laser-annealed spot on an anatase film (1 second at $\sim$185 mW, post-anneal 400 $^\circ$C for 1 hour in 20\% O$_2$) showing grain growth in the center. (b) X-ray diffraction pattern produced by summing up all the detector images acquired during a 3x3 $\mu$m$^2$ raster scan on a film patterned with Cr alignment markers. The angular $2\theta$ range covers both the anatase 101 and the rutile 110 reflections. The continuous ring along the $\gamma$ direction is formed by diffraction from crystals with random orientations. (c) Composite image showing the locations of the rutile phase. The real-space image was produced by summing up dark field images of the rutile 110 reflection acquired over a range of incidence angles between 6.0 and 12.8$^\circ$. The spots are wider in the X direction due to the projection of the incident X-ray beam at shallow incident angles. Red circles highlight three laser-exposed spots ($\sim$185 mW, 1 sec, 10 sec, 100 sec left to right). Also shown is X-ray fluorescence signal from the nearby Cr alignment structures. (d) Statistics of measured individual rutile grains extracted from spatial scans. The teal dashed line is a Gaussian fit to the data, indicating an average grain size of $88\pm 30$ nm.  }
\label{fig2} 
\end{figure*}

We observe that laser annealing at high powers causes a morphological change in the films. Figure \ref{fig2}(a) shows a scanning electron micrograph of a laser-annealed spot (1 second, 185 mW) after a 400$^\circ$C post-anneal. We see that for these conditions, the visibly small grains of the surrounding polycrystalline film undergo grain growth within the irradiated spot, which is on the order of 450 nm in diameter and is only slightly larger than the expected diffraction-limited $1/e$ Gaussian beam diameter of $0.62\lambda/$NA$\approx415$ nm for an overfilled objective lens. 

To investigate local changes in the crystal phase, we performed scanning X-ray diffraction microscopy measurements on the Hard X-ray Nanoprobe Beamline operated by the Center for Nanoscale Materials at the Advanced Photon Source at Argonne National Laboratory. A 10 keV X-ray beam was focused to a 20 nm spot and scanned across the surface of a laser-annealed film. In contrast to the chips with the arrays of 12 identical laser annealed spots, this chip was diced from the same wafer and patterned with Cr fiducial markers. Subsequent laser annealing was performed in an array between the markers, varying the laser power in each row and increasing the laser annealing time in each column of the array. This chip was then annealed for 1 hour at 400$^\circ$C in 20\% O$_2$ prior to the XRD measurements. As shown in Fig. \ref{fig2}, rutile X-ray diffraction was only observed for the highest laser annealing power (185 mW) used in this array. Figure~\ref{fig2}(b) shows the summed diffraction pattern from a raster scan covering an area of $3\times3~\mu$m$^2$. The anatase 101 reflection exhibits a continuous powder ring due to the random crystal orientation of the nanosized crystalline grains within the probed region. A donut-shaped diffraction spot was observed at the 2$\theta$ value for the rutile 110 reflection. The donut is an image of the focusing optics, the appearance of which indicates good crystallinity. We note that the crystal orientation of the rutile phase is also random, but the diffraction spots of the rutile phase do not form a continuous ring because statistically only the diffraction from one or two grains can be observed on the detector for a given incident angle. Diffraction from other rutile grains were observed after tilting the sample to different incident angles. Figure~\ref{fig2}(c) shows a dark field image of 29 rutile grains observed for an incident angle range of 6.0$^\circ$ to 12.8$^\circ$. The rutile phase was only found at a few locations scattered between the two Cr fiducial markers. These locations are consistent with the position of the laser annealed spots at a power of 185 mW. In this particular laser-annealed sample, the next lowest laser power used was $<100$ mW and no rutile signal was observed. The size of individual rutile grains was determined by performing refined line scans across individual grains and fitting the rutile diffraction signal with a Gaussian profile. The average size of the 29 grains is about 88 nm as shown in Fig.~\ref{fig2}(d).

% \begin{figure}[htbp!]
% \centering
% \includegraphics[width=0.4\textwidth]{figures/PLcomparison.pdf}
% \caption{\textbf{Add something like this somewhere?}  }
% \label{figopt} 
% \end{figure}

\begin{figure*}[htbp!]
\centering
\includegraphics[width=1\textwidth]{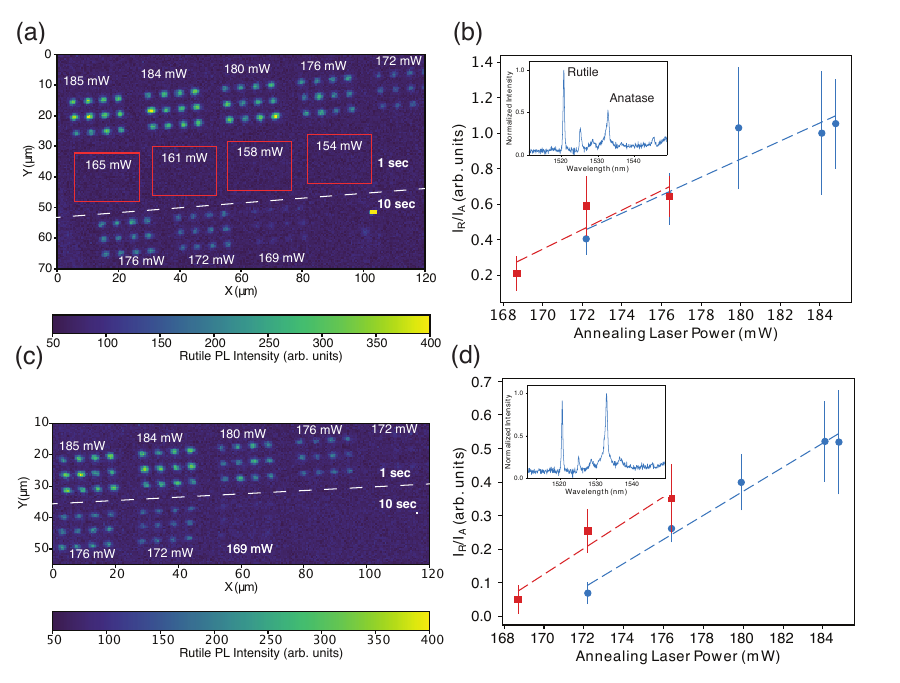}
\caption{ (a) Spatial map of the laser annealing dose array showing rutile-filtered PL. Blocks of 12 points share identical laser annealing conditions, as indicated by the labels. Regions that were annealed at lower powers that did not result in visible phase change are indicated by red boxes. (b) Rutile/anatase integrated intensity ratio as a function of laser annealing power. The points represent the average measured value for the 12 annealed spots in each array block in (a). Blue circles correspond to 1 second laser annealing times and red squares represent 10 second laser anneals. The dashed lines are linear fits to the measurement data. The inset shows the averaged spectrum for 1 second laser anneals at the maximum laser power in each case. Error bars are calculated from the standard deviation of the 12 measurements for each condition at the 95\% confidence level. Both (a) and (b) show results for the T=400 $^{\circ}$C post-annealed film. (c) The PL map and (d) integrated rutile/anatase intensity ratio for the T=500 $^{\circ}$C post-annealed film. The bright spot visible in (a) is an artifact from a scratch on the film surface. }
\label{fig3} 
\end{figure*}

We additionally investigated the effect of laser annealing power and duration on the Er:rutile transition through cryogenic PL measurements. After the laser annealing and a subsequent tube furnace annealing process, the films were cooled to T=3.6 K and excited with an off-resonant $\lambda=908$ nm laser. The PL spectra collected from the laser annealed areas showed an enhanced PL signal at about 1520 nm along with the peak at about 1533 nm, attributed to Er:rutile and Er:anatase, respectively (Fig. \ref{fig1}(b)). The rutile PL signal was selectively collected using a tunable bandpass filter set to a center wavelength of 1520 nm and a FWHM of 950 pm positioned before the superconducting nanowire single photon detector (SNSPD). As the excitation laser was scanned across the sample, clear and bright micron-scale spots were observed as in Fig. \ref{fig3}(a). We note that laser annealing powers below about 169 mW incident on the sample with 1 second duration yielded no detectable phase change while a weak rutile signal appeared when increasing the duration to 10 seconds. As shown in Fig. \ref{fig3}(c), post-annealing at 500 $^{\circ}$C seemed to have a similar effect on the PL scans compared to 400 $^{\circ}$C. While the post-anneal in the tube furnace was not critical to observe the anatase$\rightarrow$rutile transition, we ultimately found that the intensity of the principle rutile PL line was dramatically increased after the furnace annealing. We additionally attempted laser annealing with the same conditions in a partial vacuum environment (few mTorr) at room temperature and at 3.6 K; however, no rutile PL signal was observed. This is in contrast to a prior work, which observed anatase$\rightarrow$rutile transition in thin films after annealing in vacuum with high-power 532 nm excitation focused to a spot size of about 100 $\mu$m.\cite{ahmed_localized_2021} Meanwhile, the presence of oxygen vacancies and their enhanced ionic mobility in the presence of an oxygen-rich atmosphere has been proposed as a mechanism by which the anatase$\rightarrow$rutile transition can occur during photoexcitation at below-gap energies ($<$3.2 eV).\cite{benavides_laser_2018} We attribute the apparent lack of phase conversion to the reduced oxygen concentration during the laser annealing process. This further suggests a phase change mechanism that is not based on heating alone from the below-gap excitation.

Next, we remove the tunable bandpass filter and directly collect the PL signal with a spectrometer at the locations that showed the bright spots in the rutile-filtered spatial maps. Fitting each of these peaks to Lorentzian profiles, we examine the rutile/anatase integrated intensity ratio$-$in order to account for any drifts in focus$-$as the annealing laser power is increased. Figures \ref{fig3}(b) and (d) show the intensity ratio $I_R/I_A$ for films that underwent 400 $^{\circ}$C and 500 $^{\circ}$C post-anneals, respectively, while the blue circles correspond to 1 second laser anneals and the red squares correspond to 10 second laser anneals. We find that the intensity ratio increases linearly with laser power; however, for the 169 mW annealing case, phase conversion was only observed for exposure times $\geq$10 seconds. Additionally, as the post-anneal temperature increases, we observe an increase in the relative anatase intensity and a decrease in the intensity of the background PL signal. This increase in anatase signal with increasing post-anneal temperature could be potentially attributed to further crystallization and grain growth of nearby anatase grains while staying below the point of the rutile transition. It should be additionally noted that the diffraction-limited probe laser spot (at $\lambda=908$ nm) used for PL measurements would be considerably larger than the $\lambda=449$ nm annealing laser. Thus, the PL signal will include anatase phase TiO$_2$ from the periphery of the annealed spot. The decrease in the background intensity, meanwhile, is likely attributable to healing of other crystalline defects and desorption/combustion of carbon compounds on the surface that may contribute to fluorescence. 

\begin{figure}[htbp!]
\centering
\includegraphics[width=0.5\textwidth]{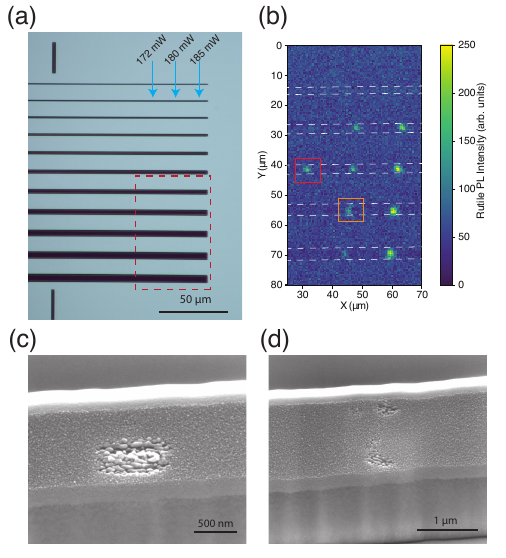}
\caption{(a) An optical micrograph of the etched ridge structures, depicting the path of the laser at the three highest powers (blue arrows). (b) The PL map corresponding to the red outlined rectangle in (a) with white dashed lines highlighting the ridge structures for clarity. (c) and (d) Scanning electron micrographs of the annealing conditions highlighted in (b) by red and orange squares, respectively. }
\label{fig4} 
\end{figure}

Lastly, we demonstrate laser-induced anatase$\rightarrow$rutile phase change on microstructures etched into the Si substrate. We first patterned ridge-like structures into the Si through standard photolithography and plasma etching. These structures range from approximately 500 nm to a couple of microns in width and are approximately 3 $\mu$m tall (Fig. \ref{fig4}(a)). After fabrication, we laser annealed the TiO$_2$ films on top of the ridges; however, instead of spot annealing, we scanned the laser across the ridges at a rate of 5 mm/sec. For each vertical cross-cut, we set a different laser power. After a 400 $^{\circ}$C post-anneal, rutile PL is visible at low temperatures for regions exposed to the three highest annealing powers, as shown in Fig. \ref{fig4}(b). Examining these annealed spots under SEM, we observe similar morphological changes in the TiO$_2$ films, often concentrated near the edges of ridge (Fig. \ref{fig4}(c-d)). Additionally, we observe occasional straggle along the direction of the ridge, which is transverse to the direction of laser scanning. We attribute this straggle to vibrations in the assembly holding the objective lens. Future improvements to the laser scanning assembly design can mitigate unwanted vibrations or straggle of the focused laser spot, which could, for example, ultimately facilitate the placement of Er:rutile at the anti-node of a photonic crystal cavity. Beyond the anatase$\rightarrow$rutile transition, sub-bandgap laser annealing has been shown to crystallize amorphous TiO$_2$ films,\cite{benavides_laser_2018} as well as to effectively transform rutile back to anatase through amorphization of the rutile phase and subsequent anatase crystallization.\cite{ahmed_localized_2021} Fully reversible phase control would provide an important tuning knob, especially when considering Er:TiO$_2$ films that are evanescently coupled to proximal photonic cavities whose resonant wavelength is deeply affected by the local environment. 

% We believe that control of the local TiO$_2$ crystal phase, and thus control of the Er$^{3+}$ optical transition 

\section{\label{sec:Conclusion}Conclusion}
In conclusion, we have demonstrated local, submicron anatase$\rightarrow$rutile phase conversion of  Er:TiO$_2$ thin films grown on Si using below-gap ($\lambda=$449 nm) laser excitation. This phase change was found to occur most readily when the laser annealing was performed in standard atmosphere and at room temperature, and the Er$^{3+}$ photoluminescence, especially that of the anatase phase, was improved by post-annealing in an oxygen-rich environment. While the rutile PL signal increased linearly with laser power, no phase change was observed by PL for annealing powers below about 169 mW, except for longer duration anneals (10 seconds). In addition, nanoscale X-ray diffraction measurements highlight the improved crystallinity of the rutile grains, vis-\`{a}-vis anatase, attributed to the laser annealing. Fine spatial scans indicated an average grain size of $88\pm30$ nm. Finally, we demonstrated that this same phase change process can extended to TiO$_2$ films on Si microstructures, which are similar in form to waveguide structures in Si photonics. Through further tuning of the anneal parameters and improvements in the laser positioning capabilities, the ability to adjust the phase of host materials with submicron resolution coupled with delta doping, opens up a new avenue for deterministic placement of quantum emitters and spin qubits.

% \section{\label{sec:Methods}Methods}

\begin{acknowledgments}
The authors thank Nazar Delegan and Robert Pettit for helpful discussions. Work at Argonne (M.V.H., D.D.A., S.G., F. J. H.) was supported by the U.S. Department of Energy, Office of Science, National Quantum Information Science Research Centers as part of Q-NEXT.  We acknowledge additional support from the U.S. Department of Energy, Office of Basic Energy Sciences, Materials Science and Engineering Division (J A.). Work by S.E.S. and M.K.S. was carried out at Argonne National Laboratory with support from the U.S. Department of Energy Office of Science Advanced Scientific Computing Research program under CRADA A22112 through the Chain Reaction Innovations program. Work performed at the Center for Nanoscale Materials and Advanced Photon Source, both U.S. Department of Energy Office of Science User Facilities, was supported by the U.S. DOE, Office of Basic Energy Sciences, under Contract No. DE-AC02-06CH11357. 
\end{acknowledgments}

% \appendix

% \section{Appendixes}

\bibliography{references}% Produces the bibliography via BibTeX.

\end{document}